\documentstyle[epsf,e-e-ijmpa,citepunct]{article}

\newtheorem{fig}[figure]{Fig.}
\begin{document} 
\ \vspace{-1.5cm}\\ LMU 01/00\hfill\vspace{0.8cm}

\begin{center}
{\bf CONSTRAINTS ON EXTRA GAUGE BOSONS IN \boldmath$e\gamma$ COLLISIONS}
\vspace{1cm}\\
STEPHEN  GODFREY, PAT KALYNIAK, BASIM KAMAL\smallskip\\
{\it Phys. Dept., Carleton Univ., Ottawa, Canada K1S 5B6,\\
E-mail: godfrey@physics.carleton.ca, kalyniak@physics.carleton.ca,
bkamal@physics.carleton.ca}\vspace{0.5cm}

ARND LEIKE\smallskip\\
{\it Ludwigs--Maximilians-Universit\"at, Sektion Physik, Theresienstr. 37,\\
D-80333 M\"unchen, Germany\\
E-mail: leike@theorie.physik.uni-muenchen.de}
\vspace{1cm}\end{center}

\begin{abstract}
We investigate the sensitivity of $e^-\gamma\rightarrow\nu_e\bar\nu_\mu\mu^-$
to extra charged gauge bosons.
The sensitivity is much below that of  
$e^-e^+\rightarrow\nu\bar\nu\gamma$.
We conclude that $e^-\gamma\rightarrow d\bar u\nu_e$ and
$e^-\gamma\rightarrow f\bar fe^-$ are also inferior to $e^+e^-$ collisions in
setting bounds on extra charged and neutral gauge bosons and on four fermion
contact interactions.
\end{abstract}

%
\section{Motivation}

Extra gauge bosons are predicted in many extensions of the Standard Model (SM).
Constraints on these particles or their discovery would constrain theories 
extending the SM. 
Therefore, the search for extra gauge bosons is foreseen in the research 
program of every future collider.

In $e^+e^-$ collisions, the process $e^+e^-\rightarrow \nu\bar\nu\gamma$ 
can constrain extra neutral ($Z'$) and extra charged ($W'$) gauge bosons 
together, 
while the processes $e^-\gamma\rightarrow d\bar u\nu_e$
and  $e^-\gamma\rightarrow\nu_e\bar\nu_\mu\mu^-$ in $e\gamma$ collisions
could constrain extra charged gauge bosons independent of neutral gauge bosons.

We focus here on the process  $e^-\gamma\rightarrow\nu_e\bar\nu_\mu\mu^-$
because it has a clean signature and is free of QCD backgrounds.
It is known to give interesting constraints on anomalous
$W$ couplings \cite{godfrey}.

Constraints on extra gauge bosons from $e^-\gamma$ collisions have to compete
with present and future constraints from other experiments, especially with
those from $e^+e^-$ collisions.
It has been shown \cite{hewett,nunug,nunugprd} that the 
process $e^+e^-\rightarrow\nu\bar\nu\gamma$ can put limits on extra charged 
gauge bosons.
For $\sqrt{s}=0.5\,TeV$, $L_{int}=500\,fb^{-1}$ and neglecting systematic
errors, one is sensitive \cite{nunugprd} to a $W'$ with a mass of 
$1.2\,TeV$--$4.6\,TeV$ depending on the model.
These limits degrade to $0.6\,TeV$--$1.8\,TeV$ if a 
systematic error of 2\% is included for the observed cross sections.

%
\section{Calculation}

%
\begin{figure}

\vspace{-4.5cm}
\mbox{\hspace{-5.5cm}\epsffile{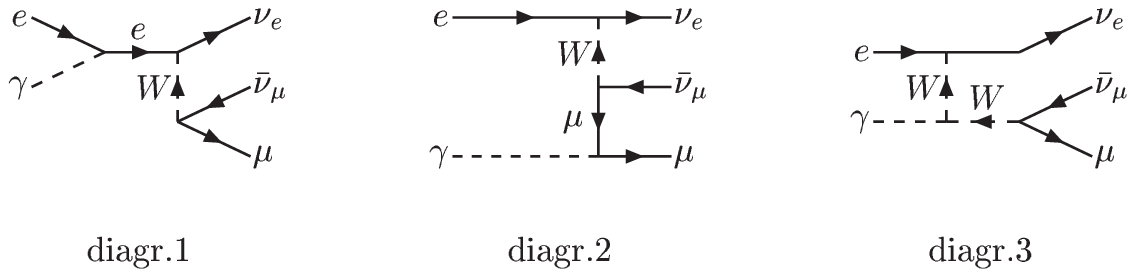}}
\vspace{-22.5cm}

\noindent
{\small\it
\begin{fig} \label{fig1} 
The lowest order Feynman diagrams for the process 
$e^-\gamma\rightarrow\nu_e\bar\nu_\mu\mu^-$.
\end{fig}}
\end{figure}

To lowest order, the process $e^-\gamma\rightarrow\mu^-\nu_e\bar\nu_\mu$ is
described by the Feynman diagrams shown in figure \ref{fig1}.
We performed two independent calculations of the cross section.
In one calculation, the squared matrix element is obtained by usual trace 
techniques with the help of the symbolic manipulation program 
{\tt form} \cite{FORM},
and integrated analytically over the phase space.
The remaining integrations are done by an adaptive Simpson routine.
In the other calculation, helicity amplitudes are calculated by spinor 
techniques \cite{BerGiele}  and then squared analytically.
The integration over the squared matrix element is performed by the MC method.
Both calculations agreed with CompHEP \cite{comphep} for the SM with
unpolarized beams.
They also agreed with each other for extensions of the SM.

For completeness, we present here the formula for the squared amplitude.
The generalized couplings may be inferred from
the $W_i l\nu$ vertex
\begin{equation}
\label{Wcoup}
W_i l\nu = \frac{ig}{\sqrt{2}} \gamma^\mu \left( \frac{1-\gamma_5}{2}
\,\,a_i + \frac{1+\gamma_5}{2} \,\,b_i \right) .
\end{equation}
We have $a_1=1$, and $b_1=0$ in the SM.

In order to present the result for the squared amplitude,
$|{\cal M}(\lambda_{e},\lambda_{\gamma})|^2$
(dependent on the polarizations 
$\lambda_{e}$ and $\lambda_{\gamma}$ of the electron and the photon),
in $e^-(p_-)+\gamma(p_+) \rightarrow \nu(q_-)+\bar{\nu}(q_+)+\mu^-(k)$,
we first define the kinematic invariants:
\begin{equation}
\begin{array}{rlrl}
\nonumber
 s =& (p_+ + p_-)^2, & s' =& (q_+ + q_-)^2, \\
\nonumber   t =& (p_+-q_+)^2, & t' = & (p_- - q_-)^2, \\
\nonumber   u = & (p_+-q_-)^2, & u' = & (p_- - q_+)^2, \\
\nonumber  k_\pm = & 2 p_\pm \cdot k, & k'_\pm = & 2 q_\pm \cdot k, \\
W_i = & k_+'-M_{W_i^2} + i M_{W_i}\Gamma_{W_i}, & 
W_i' = & t'-M_{W_i^2} + i M_{W_i}\Gamma_{W_i}.
\end{array}
\end{equation}
The resulting expression for $|{\cal M}(\lambda_{e},\lambda_{\gamma})|^2$ 
is then
\begin{eqnarray}
|{\cal M}(L,L)|^2 &=&  \frac{2(4\pi)^3\alpha^3}{s_W^4s'sk_+}
 \sum_{\stackrel{\scriptstyle i=1,n}{j=i,n}} W^+_{ij}\cdot 
(a_i^2 a_j^2 {k'_-}^2 + a_i a_j b_i b_j {s'}^2) \\
|{\cal M}(L,R)|^2 &=&  \frac{2(4\pi)^3\alpha^3}{s_W^4s'sk_+}
 \sum_{\stackrel{\scriptstyle i=1,n}{j=i,n}} W^-_{ij}\cdot 
(a_i^2 a_j^2 u^2 + a_i a_j b_i b_j k_-^2) \\
|{\cal M}(-\lambda_{e},-\lambda_{\gamma})|^2 &=& 
|{\cal M}(\lambda_{e},\lambda_{\gamma})|^2 [a\leftrightarrow b,
W^\pm_{ij} \leftrightarrow W^\mp_{ij}]
\end{eqnarray}
with
\begin{equation}
W^\pm_{ij} = (2-\delta_{ij}) {\rm Re}(F^\pm_i {F^\pm_j}^*)
\end{equation}
and
\begin{equation}
F^\pm_i = {\displaystyle \frac{s'}{W_i} + \frac{tt' + ss' - uu' 
\pm 4i\varepsilon_{\mu\nu\rho\sigma}p_-^\mu p_+^\nu q_-^\rho q_+^\sigma}
{2 W_i W'_i}}, 
\end{equation}
where $\varepsilon_{\mu\nu\rho\sigma}$ is the completely antisymmetric
Levi-Civita tensor and $\varepsilon_{0123}=1$. The summation runs
over the number, $n$, of charged gauge bosons in the theory ($n=1$ in the SM
and $n=2$ in a theory with one $W'$).

For numerical calculations, we use the physical input
$M_W=80.33\,GeV$,\\
$\Gamma_W=2.06\,GeV$, $\alpha=1/128$, $\sin^2\theta_W=0.23124$ and $m_\mu=0$.

The massless muon does not lead to a divergent cross section because we 
apply the following cuts on the angle between the muon and the photon and 
on the muon energy,
\begin{equation}
10^\circ < \theta(\gamma,\mu) < 170^\circ,\ \ \   E_\mu > 10\,GeV,
\end{equation}
to ensure that the muon gives a signal in the detector.
We assume 90\% polarization of the electron and photon beams.

As observables, we consider different polarized and unpolarized cross sections,
\begin{equation}
\sigma,\ \sigma_{LU},\ \sigma_{RU},\ \sigma_{LL},\ \sigma_{LR},
\ \sigma_{RL},\ \sigma_{RR}.
\end{equation}
The first (second) index refers to the electron (photon) helicity.
In addition, we consider the distributions $d\sigma/d c$, 
$c=\cos\theta(\gamma,\mu)$ and $d\sigma/d E_\mu$.
We assume a systematic error of 2\% in these observables.
Therefore, the $W'$ must produce in these observables a deviation of at 
least 2\% to give a signal.

We do not convolute over the photon energy spectrum here.
We understand that this would be needed in a final calculation of the 
process.
However, it would not change our main conclusions.

We present numerical results for different models containing extra charged
gauge bosons.
See reference \cite{nunugprd} for a more detailed description of 
the models.
Here is a short summary to make this paper self contained.

{\bf LRM:} Left-right-symmetric model with $\kappa=g_R/g_L$, $\rho=1 (2)$
for symmetry breaking through Higgs doublets (triplets).

{\bf SSM:} Sequential SM, the $W'$ is a heavy repetition of the SM $W$,
$a_2=1,\ b_2=0$.

{\bf SSM $W'_R$:} As the SSM, but with a right-handed $W'$, $a_2=0,\ b_2=1$.

{\bf UUM:} Un-unified model, quarks and leptons are gauged by different
$SU(2)$ gauge groups, $G=SU(3)_c\times SU(2)_q\times SU(2)_l\times U(1)_Y$,
characterized by a mixing angle, $\phi$, which  
represents the mixing between the charged bosons of 
the two $SU(2)$ symmetries.

{\bf KK:} Kaluza Klein excitations, here only the first excitation is 
considered.
The corresponding $W'$ couples as in the SSM but with fermionic couplings 
which are a factor $\sqrt{2}$ larger. The $\gamma W W$  coupling is unaffected
since there is only the SM photon, which couples to charge as usual.

%
\section{Results and Discussion}

\begin{figure}[tbh]
\ \vspace{3cm}\\
\begin{minipage}[t]{6.0cm} {
\begin{center}
\hspace{-0.8cm}
\mbox{
\epsfysize=6.2cm
\epsffile[0 0 500 500]{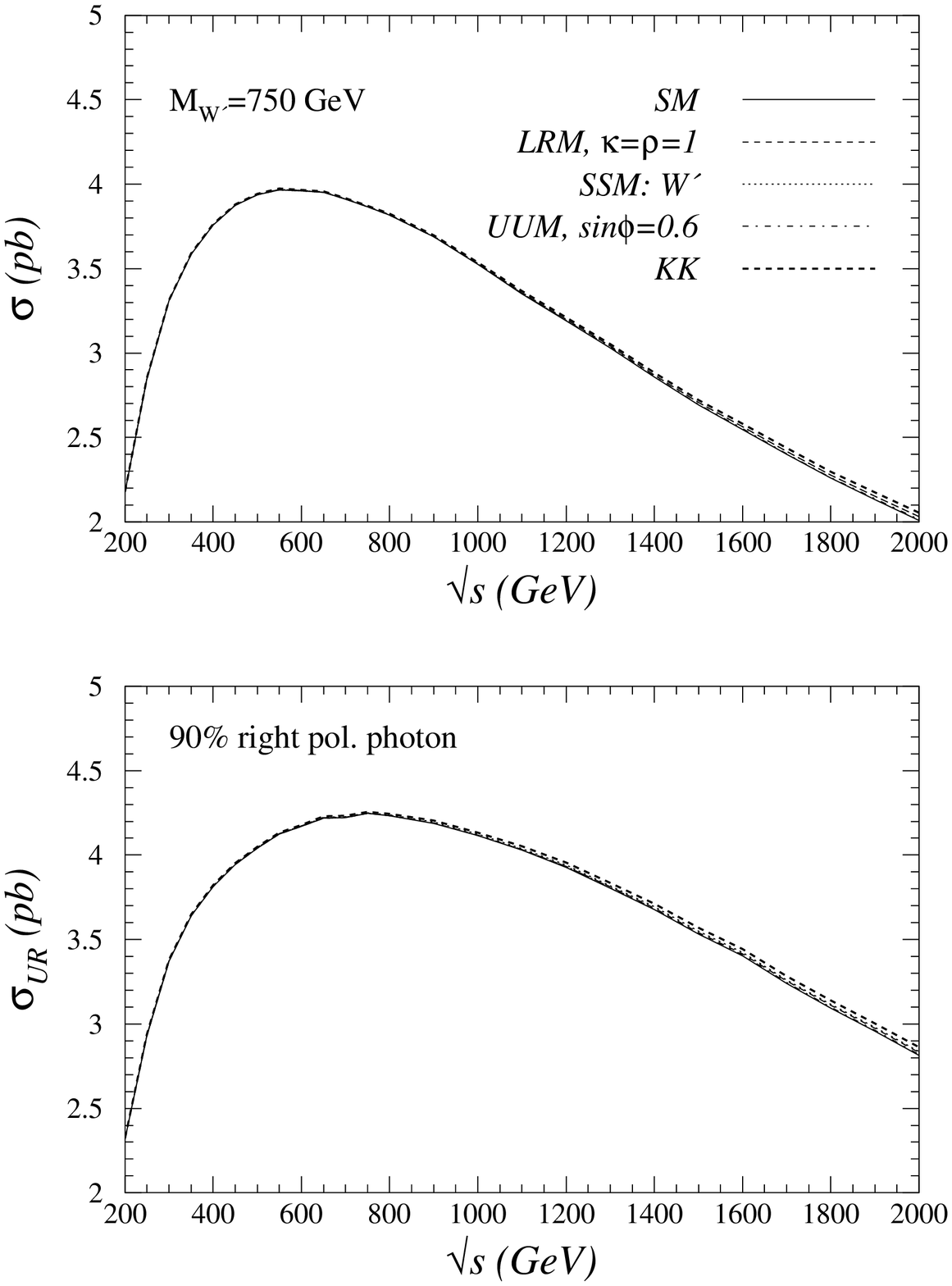}
}
\end{center}
}\end{minipage}
\hspace*{0.5cm}
\begin{minipage}[t]{6.0cm} {
\begin{center}
\hspace{-1.0cm}
\mbox{
\epsfysize=6.2cm
\epsffile[0 0 500 500]{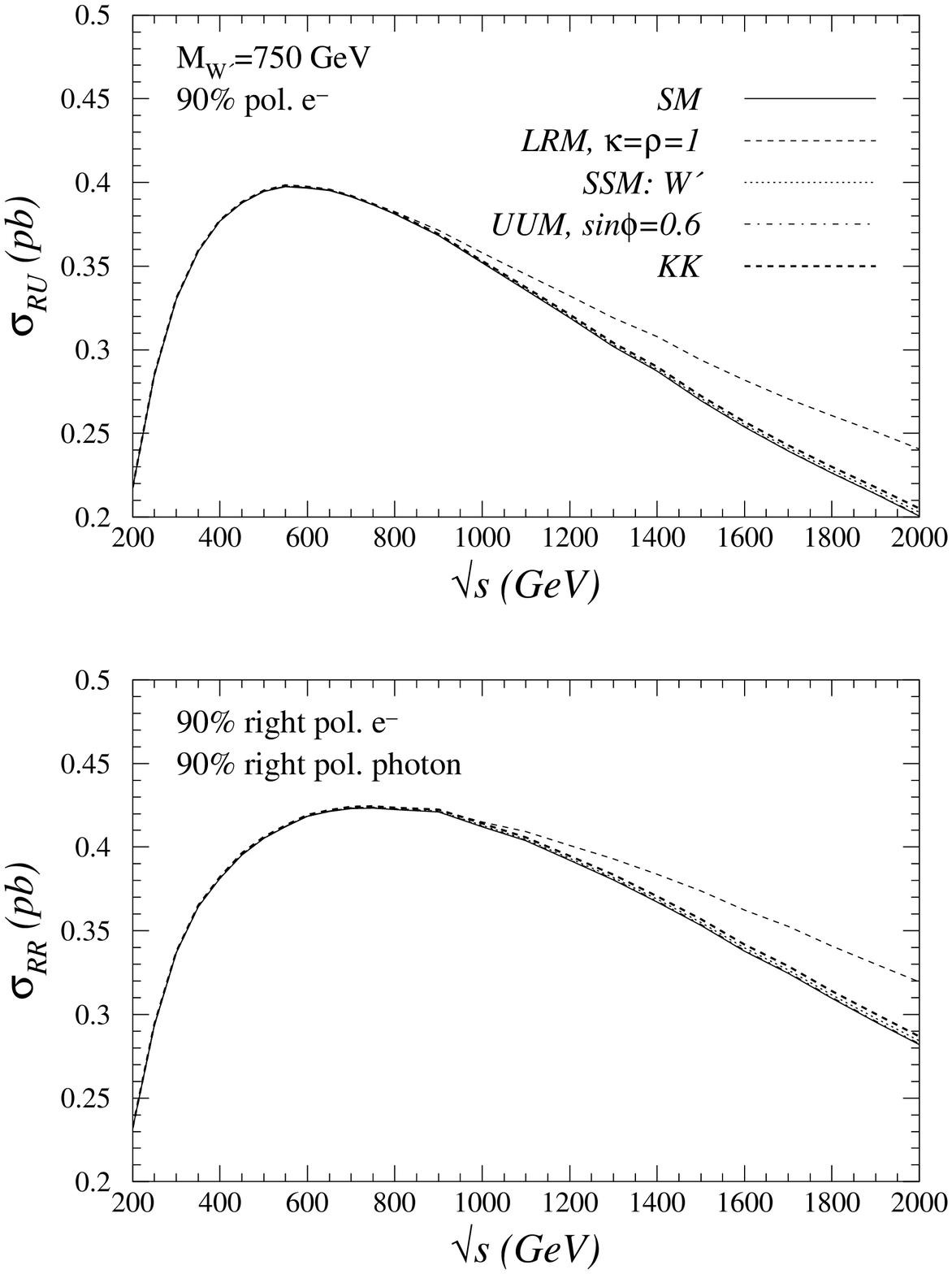}
}
\end{center}
}\end{minipage}
\vspace*{-0.5cm}
\noindent
{\small\it
\begin{fig} \label{sgeuvs} 
The total crosss sections $\sigma,\ \sigma_{RU},\ \sigma_{UR}$ and 
$\sigma_{RR}$ as a function of $\sqrt{s}$ for $M_{W'}=750\,GeV$.
If polarized, 90\% polarization of the electron and photon beams is assumed.
Results are given for the SM (solid line), LRM (dashed line), 
SSM (dotted line), UUM (dash-dotted line) 
and the KK model (thick dashed line).
\end{fig}}
\end{figure}

Figure 2 shows total cross sections for different beam polarizations and for
different models.
We see that the process has a cross section of several $pb$.
Unfortunately, the deviation from the SM is very small for the considered 
models.
It can be enhanced by right-handed polarized electrons and photons.
However, the process  $e^-\gamma\rightarrow\nu_e\bar\nu_\mu\mu^-$
remains much less sensitive to a $W'$ than the process
$e^-e^+\rightarrow\nu\bar\nu\gamma$.
Sensitivity to models with a right-handed $W'$ could be enhanced with higher
degrees of polarization, of course, since the SM contribution vanishes
for 100\% right-polarized electrons.
Left-handed electrons or photons give a sensitivity which is not better 
than that for unpolarized beams.

\begin{figure}[tbh]
\begin{minipage}[t]{6.0cm} {
\begin{center}
\hspace{-0.8cm}
\mbox{
\epsfysize=5.5cm
\epsffile[0 0 500 500]{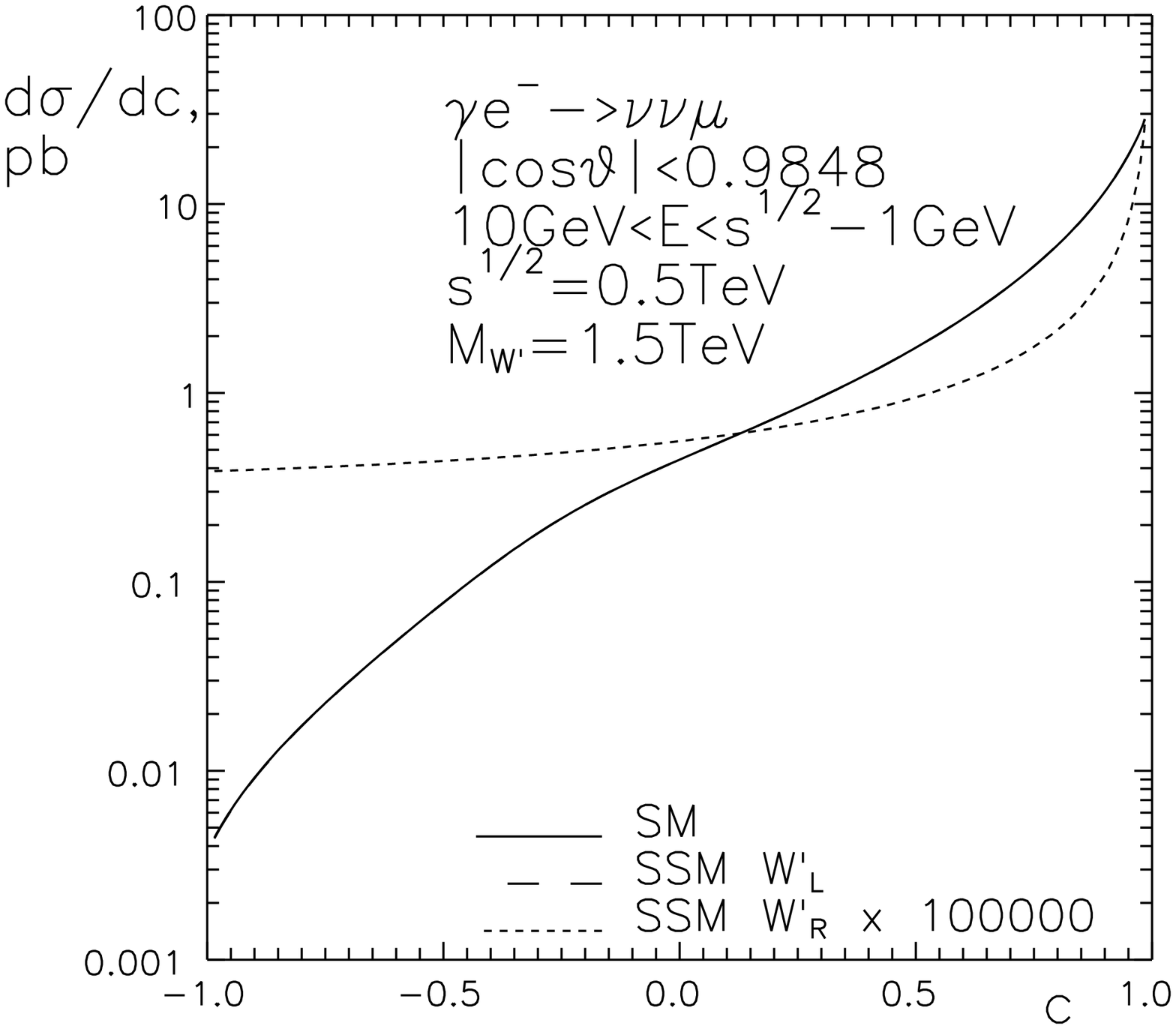}
}
\end{center}
\noindent
{\small\it
\begin{fig} \label{nnm3} 
The angular distribution $d\sigma/d c$ 
for $\sqrt{s}=0.5\,TeV$ and $M_{W'}=1.5\,TeV$.
Shown are the results for the SM (solid line), the SM with a left $W'$ 
(long dashed line) and a right $W'$ alone (short dashed line).
\end{fig}}
}\end{minipage}
\hspace*{0.5cm}
\begin{minipage}[t]{6.0cm} {
\begin{center}
\hspace{-1.0cm}
\mbox{
\epsfysize=5.5cm
\epsffile[0 0 500 500]{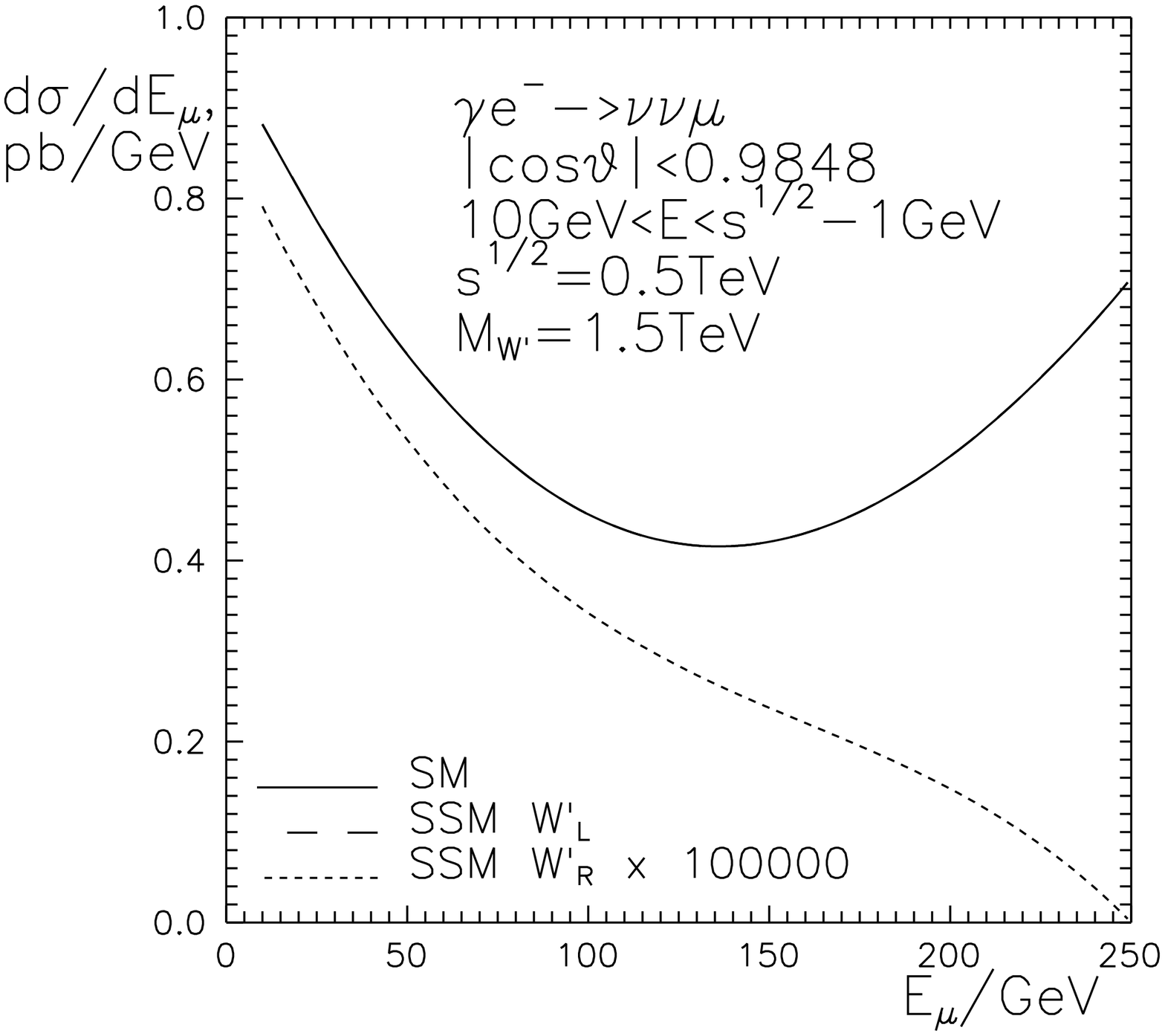}
}
\end{center}
\noindent
{\small\it
\begin{fig} \label{nnm2} 
The Energy distribution $d\sigma/dE$ for $\sqrt{s}=0.5\,TeV$ and 
$M_{W'}=1.5\,TeV$.
Shown are the results for the SM (solid line), the SM with a left $W'$ 
(long dashed line) and a right $W'$ alone (short dashed line).
\end{fig}}
}\end{minipage}
\end{figure}

Figures~\ref{nnm3} and \ref{nnm2} show the angular and energy distributions of 
the muon for different models.
For the SSM $W'_R$, only the $W'_R$ contribution is shown, for illustration.
Compared to the total cross sections, the distributions give no enhancement 
in the sensitivity.
(A lighter $W'$ would give a larger signal.)
A $W'$ of 1.5\,TeV {\em does} produce a signal in  
$e^+e^-\rightarrow \nu\bar\nu\gamma$ \cite{nunugprd}.
In our process however, the SM and SSM lines are indistinguishable.
Their difference is much below a reasonable systematic error of about 2\%.
The prediction for the SSM $W'_R$ is different from the SM but it is suppressed
by a factor of more than $10^5$ as there is no interference with the SM. 
Even in the case of 100\% right-polarized electrons, there
would be only a few events.
For realistic degrees of polarization, these events are contaminated by
hundreds of thousands of SM events coming from left-handed electrons.

We see from figure~\ref{nnm3} that the muonic angular distribution is 
strongly peaked in the direction of the photon (note the 
logarithmic scale).
This indicates that diagram  2 in figure~\ref{fig1},
where the muon is exchanged in the $t$ channel, dominates the process.
In this diagram, the charged gauge boson is also exchanged in the $t$ channel.
The corresponding propagator is largest for the $\nu_e$ travelling down the 
beam pipe.
In this case, the propagator can be simply replaced by $1/M_W^2$ or 
$1/M_{W'}^2$.
The exchange of the light SM $W$ gives many events which contaminate a
potential $W'$ signal.
In rough approximation, the fractional deviation from
the SM cross section due to a left-handed $W'$ is
\begin{equation}
\label{estim}
[\sigma(SM+W') - \sigma(SM)]/\sigma(SM) \approx M_W^2/M_{W'}^2.
\end{equation}
For $M_{W'}=1.5\,TeV$, this leads to a deviation of 0.3\%.
For a right-handed $W'$, the deviation is proportional to the square of
(\ref{estim}), as there is no interference with the SM. Consequently,
the sensitivity to $W'$'s below threshold is rather poor.
As we go to higher energy, the sensitivity increases due to the 
larger phase space for neutrinos, which are not peaked along the beam pipe.
Hence there will be an additional $s$-dependence in the deviation 
estimated in equation (\ref{estim}). 

Let us remark about why 
the process $e^+e^-\rightarrow\nu\bar\nu\gamma$
is much more sensitive to a $W'$ than the process 
$e^-\gamma\rightarrow\nu_e\bar\nu_\mu\mu^-$.
There are two clear reasons why  $e^+e^-\rightarrow\nu\bar\nu\gamma$ is
more sensitive. First, there is an interference between the $W'$ and
the SM $Z$ for right-handed (and left-handed) $W'$'s. Second, the majority
of the events come from regions in the angular phase space where
we are most sensitive to a $W'$. In 
$e^-\gamma\rightarrow\nu_e\bar\nu_\mu\mu^-$, the opposite is true as
can be seen from figure 3. Even in the region where the relative deviation 
from the SM is largest, the effect of the left-handed $W'$ is quite small
and that of the right-handed $W'$ is negligible (because the huge SM 
background contaminates a potential signal).


Similar arguments could be repeated for the processes
$e^-\gamma\rightarrow d\bar u\nu_e$ and $e^-\gamma\rightarrow f\bar fe^-$,
leading to the conclusion that these processes are also much less 
sensitive to a $W'$ or $Z'$ than 
$e^+e^-\rightarrow\nu\bar\nu\gamma$ or $e^+e^-\rightarrow f\bar f$.
We finally conclude that the considered processes
in $e^-\gamma$ collisions are  not competitive to those in $e^+e^-$ collisions 
in putting limits on four fermion contact interactions either.

\end{document}